# Real-time Recognition of Smartphone User Behavior Based on Prophet Algorithms

Mi Chunmin[1], Xu Runjie[2*], Ching-Torng Lin[3]

[1] College of Economics and Management, Nanjing University of Aeronautics and Astronautics, 210016, China
cmmi@nuaa.edu.cn

[2] College of Economics and Management, Nanjing University of Aeronautics and Astronautics, 210016, China
Corresponding author: lilipota@163.com

[3] Da-yeh University, Department of IT, Changhua, 51591 Taiwan
charllin@mail.dyu.edu.tw

*Abstract:*Smartphones have become an important tool for people's daily lives, work and learning. This brings higher requirements for the security of mobile terminals such as smart phones, especially in mobile shopping, mobile payment and other application scenarios. Although the traditional physical password, fingerprint unlocking and facial features have improved the security to a certain extent, they have the characteristics of passive authentication and easiness to be stolen. The existing behavioral data collected based on mobile phone sensors is mainly used for human activity recognition and fall detection and health management. Prophet is a procedure for forecasting time series data based on an additive model where non-linear trends are fit with yearly, weekly, and daily seasonality, plus holiday effects. It works best with time series that have strong seasonal effects and several seasons of historical data. Prophet is robust to missing data and shifts in the trend, and typically handles outliers well. Based on the time series behavior data of mobile terminal users, this paper uses Prophet algorithm to decompose the time series of six kinds of daily behavior and strip off the singular value, to get the inherent cycle and trend of each behavior, and to verify the legitimacy of the behavior user at the next moment. The experimental results on the UniMiB SHAR public dataset show that the user only needs to do 2 cycles of specified actions to realize the prediction of the next time series. The main contribution of this paper is that we propose a new idea for smartphone user authentication. It is based on real-time data of smartphone user behavior, through Phophet algorithm for feature decomposition and time series prediction, and to find the inherent cycle and other characteristics, so as to perform user behavior recognition. This data-driven auxiliary authentication method can effectively solve the problem of easy forgery of static feature recognition such as password, fingerprint and face recognition.

# INTRODUCTION

The vigorous development of smartphone industry has brought a series of personalized applications (such as mobile transfer, e-mail, mobile payment, smart home, etc.) [1][2]. These applications provide people with more convenience and better user experience. However, behind the benefits of smartphones, there are often hidden risks of mobile phone information leakage. Research shows that for the PIN password or other touch passwords of smartphones, attackers can detect the position of screen click on smartphones based on acceleration sensors and gyroscopes, and then get letters, numbers or sliding gestures on the screen, and crack the PIN password. Attackers can even extract a password from the residual oil on a smartphone screen [3][4][5]. There are also studies showing that mobile phone theft and other criminal incidents occur every year in various countries, and continue to grow. The loss of these smart devices will not only cause the loss of property of the owners, but also cause the leakage of privacy data inside the mobile phone [6][7].

Therefore, there are obvious limitations in using physical passwords to protect user data security, so a non-active user authentication mechanism is very necessary, which makes biometrics become a popular research field [8][9][10][11]. At present, biometric research has achieved relatively high accuracy, but it is still difficult to be applied in mobile devices, which mainly results from dependence on hardware, and high sensitivity of physiological mechanisms to some environmental factors. For example, facial recognition requires that the sensing device of the device must be correctly aligned with the face, while speech recognition has more stringent environmental requirements for background noise, etc.

Some scholars have proposed a behavior recognition model based on mobile acceleration sensor [12][13][14]. This recognition model can operate normally without other hardware devices, and the authentication mode is transparent, persistent and difficult to imitate. Among the researches of mobile phone sensor behavior recognition, the scenarios mainly focus on the active interaction between users and smart devices, such as screen interaction, sensor interaction, application interaction [15][16]. However, if the user does not turn on the mobile screen, can he or she still recognize the behavior? For example, after the user's mobile phone is stolen, the perpetrator may not unlock it instantly, but rather choose to leave the crime scene. If we can identify the illegal carriers in time and establish a privacy protection mechanism, it will be of great significance. In addition, users have limited time to use smartphones every day, but as long as the mobile phones and other smart wearing devices are carried, users can fully record their activities in the day. Therefore, it is more meaningful to study the complex data generated by mobile phones and other wearable devices in the process of carrying than just exploring the user behavior when using smartphones.

In order to combine the above-mentioned authentication problems in non-interactive scenarios, and achieve the balance between energy consumption and authentication accuracy in the authentication process, this paper presents a framework design running in the backstage. The whole system includes three parts: behavior model, recognition mechanism and prediction model. Using smartphone sensors is a typical multi-variable time series classification problem. It uses sensor signals and extracts features from them to identify activities through classification [17]. The idea of behavior recognition provided in this paper mainly comes from the user's personal habits in daily life [18][19][20]. Unlike other scenarios which require entering password, this system does not need to turn on the screen of the mobile phone, nor trigger device interaction. It only needs to be carried by the user to judge the legitimacy of the user. Through the training of UniMiB SHAR data set by Prophet algorithm, we empirically analyzed the behavior characteristics of device carriers under six kinds of daily activities. The results show that the method can predict the confidence interval for the next moment after users make two periodic actions, and can achieve continuous time series prediction, and better verify the legitimacy of the logged-in users in the subsequent use process. The system can cooperate seamlessly with the existing password authentication mechanism in mobile devices and play the role of implicit authentication, thus providing help for more secure scenarios.

## Data and algorithm

The few open data sets of terminal sensors can be divided into two main categories: data set acquired by environmental sensors and data set acquired by wearable devices. In this paper, considering that the data set studied must satisfy the application scenario without human-machine interaction, we choose UniMiB SHAR data set [20].

Samsung Galaxy Nexus 19250 equipped with Android 5.1.1 and Bosh BMA220 acceleration sensors is used to collect the data set. A total of 11711 activities were collected from 30 subjects aged 18 to 60. The activities were divided into 9 kinds of daily activities and 8 kinds of fall tests. The data set includes three vertical axis accelerations with a constant sampling rate of 50 Hz.

The main analysis algorithm used in this paper is Propher [21]. Proposed by Facebook, this algorithm is a prediction algorithm based on time series decomposition and machine learning. It has superior performance in dealing with default values and large-scale automatic prediction.

Prophet uses Decomposition of Time Series, which includes the main conditions needed for time series data analysis: overall trends, periodicity and noise. Its basic model is:

$$y(t) = g(t) + s(t) + h(t) + \varepsilon_t$$

In which, $g(t)$ indicates the growth function to fit the non-periodic changes, $s(t)$ indicates the periodic qualitative changes, $h(t)$ indicates the special changes caused by fixed time periods (such as holidays), and $\varepsilon_t$ indicates the noise.

The growth function $g(t)$ is defined as a logical function:

$$g(t) = \frac{C}{1 + exp(-k(t-b))}$$

This function is similar to the population growth function, in which $C$ is capacity, $k$ is growth rate and $b$ is offset. With the increase of time, $g(t)$ will tend to be $C$.

The classic case of periodicity in Prophet is the periodic changes brought about by seasonality, which are approximated by Fourier series.

$$s(t) = \sum_{n=-N}^{N} c_n e^{i\frac{2\pi n}{P}}$$

In dealing with periodic problems, Prophet's task is to fit $c_n$ coefficient. The larger $N$ is, the higher the ability to fit complex cycles is.

The variable represented by $h(t)$ is very important in the analysis of time series, such as the significant changes of the number of people in shopping malls on weekends and festivals. The method Prophet uses for this purpose is to extract these dates separately and set a virtual variable.

Prophet uses the cyclic analysis method of manual regulation and automated prediction to make large-scale data prediction. It uses simulated historical predictions to evaluate performance outside the sample, and modifies predictions with problems. It has the following characteristics:

• It can process large-scale, fine-grained data and flexibly switch data resolution between micro and macro levels.

• Compared with other modeling tools, Prophet can be used without continuous time series data, so the problem of data defaults can be neglected.

• It can identify mutation points and abnormal values, and have a variety of anti-mutation and regulation methods.

# Recognition Model

According to the research on behavior recognition in relevant literature [17][20][22], we propose a behavior recognition model for smart devices in non-interactive scenarios.

Firstly, in order to accurately characterize an individual's unconscious behavior habits, the behavior model should include two characteristics: user behavior and device response. For the simple and easy repetitive actions we have studied, there are the following basic elements: whether there is periodicity in frequent repetition of actions, the singular values caused by the accuracy of sensors in measuring, and the noise caused by other factors.

The device observes various actions in a day's record, and these actions may be repeated in a day. We regard it as a system to be expressed by $O_i = \{S_i, f_{a1}, f_{b1}, f_{c1} \cdots, f_{in}\}$, in which $O_i$ represents the whole behavioral system of the user whose serial number is $i$ detected by the mobile phone sensors; $S_i$ represents a series of singular values which are inevitable when sensors record, this is because that minor changes of different gestures and different holding postures will cause difference of vibration values [3][23][22]; $f_{in}$ represents the $i$ type of actions being observed for the $n$ times.

For any action $i$, we express the behavior model of the action as $Y(t)$:

$$Y(t) = G(t) + S(t) + H(t) + \varepsilon_t$$

In which $Y(t)$ represents the prediction model of the behavior at $t$ time; $G(t)$ represents the growth function to fit the behavior without obvious periodicity; $S(t)$ represents the periodicity of the behavior; $H(t)$ represents the singular value of sensors; $\varepsilon_t$ represents the noise caused by other environmental factors.

# Recognition Strategy

In practical applications, as a safety auxiliary function for identity authentication, behavior recognition often needs high credibility of conclusions, which may need a long period of model training to provide better service to users. Although continuous observation of sensors can indeed collect abundant behavioral information for the owner's personal model training, the main energy cost of behavioral recognition is caused by sensors, and continuous observation may result in useless energy loss [22][24]. For example, when users put mobile phones on the desktop for users sitting for a long time or users playing games, it is unnecessary for the sensors to be activated continuously.

However, if the sensors are shut down for a long time, the incident may still occur in the meantime, and the risk will increase with the increase of detection stop time. This reflects the difference between identity recognition and behavior recognition: the purpose of behavior recognition is accuracy, while that of identity recognition is safety. In order to reduce energy consumption without decreasing recognition accuracy [25][26], a good observation timing mechanism still needs to be designed.

For this purpose, we design a mechanism to determine the time of starting and shutting down sensors, so as to achieve the balance between prediction accuracy and energy loss. The aim is to use the terminal equipment efficiently for the training of "master model". Of course, this assumes that the device is carried by a single user on a daily basis.

Any data will go through three steps under this mechanism: identification, judgment and retraining.

• Firstly, the system needs to recognize the action of the newly collected data. In many literatures, it can be found that models such as SVM can tag the user's motion type [22].

• Then, the new collected data are compared with the predicted results of the master model. In the following case study, we give the confidence interval of behavior recognition prediction results, which we define as "Tolerable Error (TE)".

• Finally, if there is this error range between the new data and the predicted results of the original model, the system will not be warned of potential safety hazards by adding the new data set to the later iteration process of the model; if the predicted results of the model are far from the actual measured values, or have exceeded the confidence interval range, the data acquisition intensity will be enhanced, and the legitimacy of users will be further confirmed.

## Performance Evaluation

Among UniMiB SHAR data sets, we selected six kinds of data, Walking, Running, Standing up, Sitting down, Lying down and Jumping, as the reference actions of behavior recognition. These six kinds of actions are different from the complex behaviors such as dining and outdoor mountain climbing. The selected actions in this paper are called simple actions. From our experimental process, simple behavior has higher recognition accuracy in behavior recognition and identity authentication. On the one hand, these behaviors are moderately complex and very common in life, which are very suitable as reference variables for identity authentication. On the other hand, these behaviors have their own periodicity in the process of movement, making it easy to mine features.

If we observe through the three vertical acceleration data in the data set directly, we find that the periodicity of each action cannot be obtained directly, and it is difficult to define the behavior recognition model of the subject. Unlike previous research methods to achieve behavior recognition and user authentication, we use time series prediction method to predict the next set of time series through training data sets, and then use prediction to obtain confidence intervals to be compared with new data, so as to achieve the purpose of judging legitimate users and intruders.

In this experiment, we use Prophet algorithm to identify mutation points and abnormal values. As shown in Figure 1 below, black represents the discrete points of the original time series, and dark blue lines represent the values obtained by fitting the time series, while light blue region represents the confidence intervals of time series.

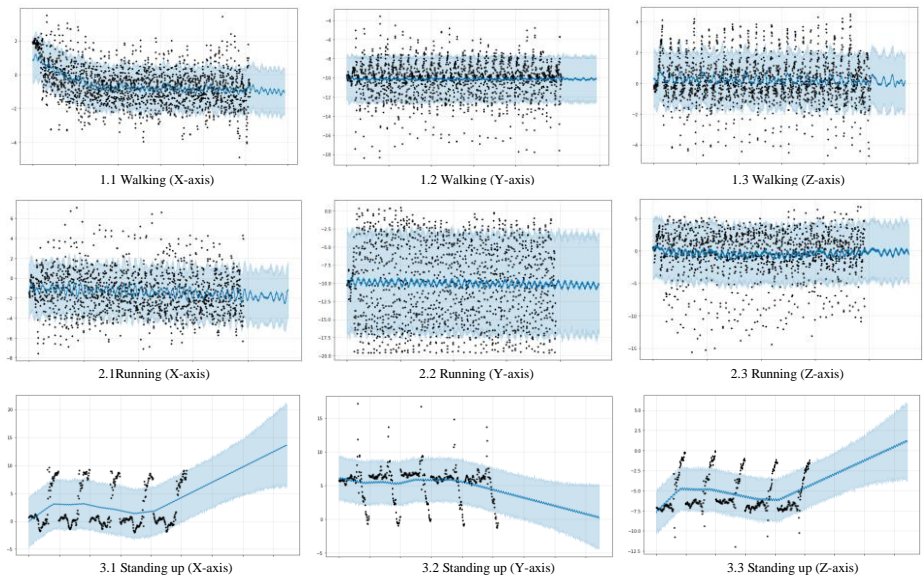

1.1 Walking (X-axis)　　1.2 Walking (Y-axis)　　1.3 Walking (Z-axis)

2.1 Running (X-axis)　　2.2 Running (Y-axis)　　2.3 Running (Z-axis)

3.1 Standing up (X-axis)　　3.2 Standing up (Y-axis)　　3.3 Standing up (Z-axis)

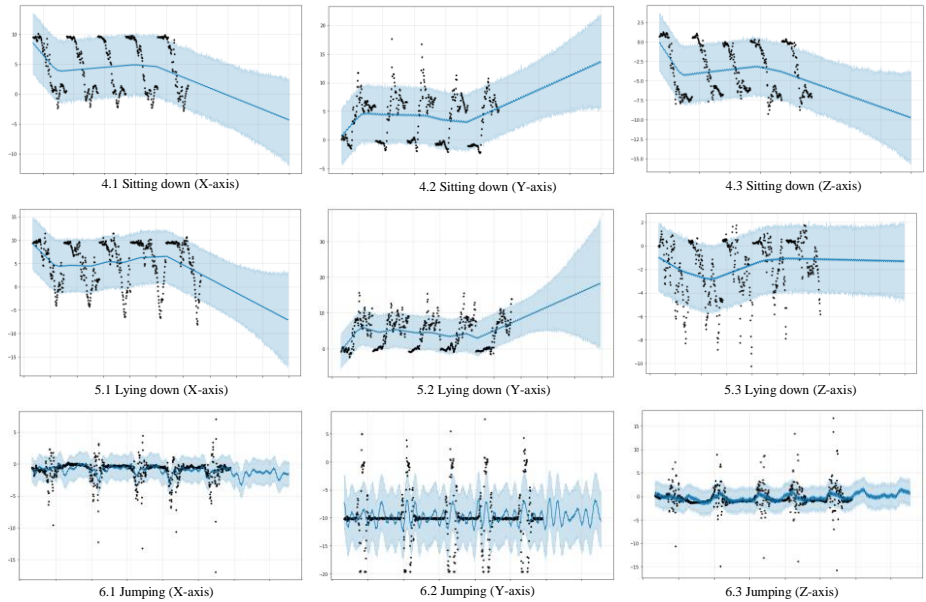

**Fig.1 Visualization of 3 Vertical Axis Accelerations**

According to the results, Prophet makes a prediction of 300 time nodes in the future. The X-axis, Y-axis and Z-axis values of the subject moved with equal amplitude around the midline respectively. The magnitude of the amplitude indicates the subject's behavioral habits in daily activities. Because of the intenseness of activities, there are singular values in the data, but the singular values are often distributed in a fixed area, which is captured by the anti-mutation ability of Prophet algorithm, so it also has a certain trend prediction in its future prediction.

In order to test the ability of Prophet algorithm in recognizing human behavior and realizing identity authentication, we cross-validate the experiment. The first 500 time series of Jumping data are taken as training set, and then the prediction is made every 100 groups, and the real data are used to test the prediction data. A total of five groups of cross-validation are performed, as shown in Figure 2. The results show that after the user makes the second cycle of actions, Prophet algorithm can predict the user's behavior habits well. The Prophet algorithm is effective in predicting, modeling and identifying human behavior data, and can be used as an auxiliary authentication method for user safety.

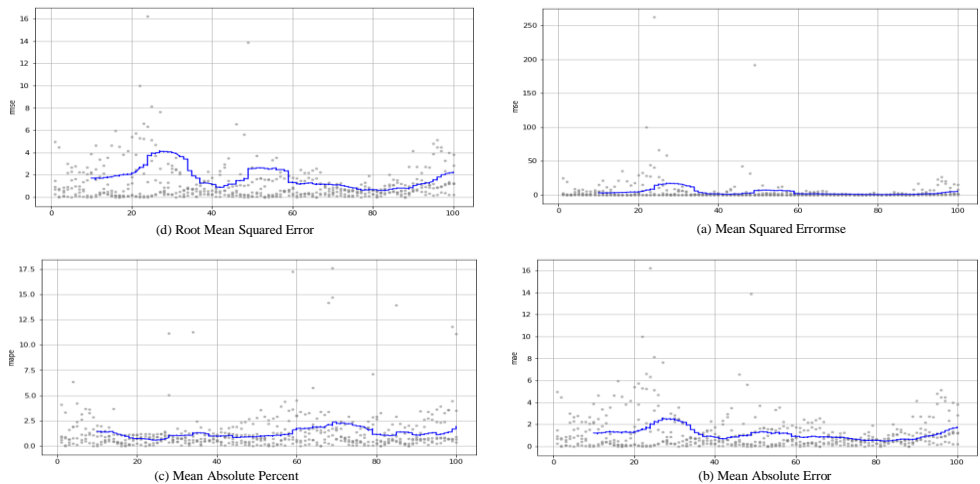

**Fig.2 Prediction Error in Cross-validation of Time Series**

## Conclusion

In this experiment, six groups of common behaviors in daily life are used as model training data for human behavior recognition. After using Prophet algorithm to model and predict six groups of behavior data, the experiment has achieved good results and proven that the behavior recognition and identity authentication system in non-human-machine interaction scenarios can be realized.

In real-life scenarios, we consider that the user's own behavior habits are not unchanged. If a fixed model is always used to predict and identify the user's behavior habits, it is not accurate enough. Only the continuous change of the model can better adapt to the owner's own behavior habits. The Prophet algorithm is very adaptable to periodic data and can be adjusted gradually according to the later changes, which is a step forward for behavior recognition. In addition, Prophet algorithm is very friendly to time series data with missing values, which means that the shutdown and placement of mobile phones will not cause the failure of the model, which makes the identification based on behavior data have a more reliable solution.

In many scenarios, the application of biometric authentication assisted by behavior data has broad prospects. Only after the owner of the device is required to carry the device for a period of time, the data accumulated by the user for two cycles of actions is enough to describe a recognition framework for the device. If an attacker maliciously cracks a physical password or uses a mobile phone to perform sensitive operations (such as transferring, trading, checking bank card information, etc.), behavior model can be used to continuously and invisibly authenticate the attacker's identity and prevent the risk of information leakage.

**Acknowledgments:** This work is supported by the National Social Science Foundation of China (17BGL055).